\documentclass{article}

\usepackage{mathptmx,float,authblk}       
 \usepackage{amsmath,mathrsfs}
 \usepackage{amssymb,latexsym,amsfonts,graphicx}

 \usepackage{color}
\usepackage{ulem}
\usepackage{makeidx}         
\usepackage{graphicx,wrapfig}        
\usepackage{multicol}        
\usepackage[bottom]{footmisc}
\usepackage[fleqn]{mathtools}

\makeindex             

\title{Whitham modulation equations  and application to small dispersion asymptotics and long time asymptotics of nonlinear dispersive equations} 
\begin{document}


\author[1,2]{Tamara Grava}
\affil[1]{SISSA, Via Bonomea 265, 34136 Trieste, Italy, grava@sissa.it }
\affil[2]{School of Mathematics, University of Bristol, Bristol BS8 1TW, UK}
%
\maketitle

\abstract{In this chapter we review the theory of modulation equations  or Whitham equations for the travelling wave solution of KdV. We then apply the Whitham modulation equations to describe the   long-time asymptotics and small dispersion asymptotics of the KdV solution.}

\section{Introduction}

The theory of modulation refers to  the idea of slowly changing the constant  parameters in a solution to a given PDE.
Let us consider for example the linear PDE in one spatial dimension
\begin{equation}
\label{linear}
u_t+\epsilon^2 u_{xxx}=0,
\end{equation}
where $\epsilon$ is a small positive parameter.
Such equation admits the exact travelling wave solution
\[
u(x,t)=a\cos \left(k\dfrac{x}{\epsilon}+\omega \dfrac{t}{\epsilon}\right),\quad \omega=k^3
\]
where $a$ and $k$ are constants.
 Here $\dfrac{x}{\epsilon}$ and $\dfrac{t}{\epsilon}$ are considered as fast variables  since $0<\epsilon\ll 1$.
The general solution of equation (\ref{linear})  restricted for simplicity to  even initial data $f(x)$   is given by
\[
u(x,t;\epsilon)=\int_0^{\infty}F(k;\epsilon)\cos \left(k\dfrac{x}{\epsilon}+\omega \dfrac{t}{\epsilon}\right)dk
\]
where the function $F(k;\epsilon)$ depends on the initial conditions by the inverse Fourier transform $F(k;\epsilon)=\dfrac{1}{\pi \epsilon}\int_{-\infty}^{\infty}f(x)e^{-ik\frac{x}{\epsilon}}dx$.

For fixed $\epsilon$,  the large time asymptotics of $u(x,t;\epsilon)$  can be obtained using the method of stationary phase
\begin{equation}
\label{expansion_u_lin}
u(x,t;\epsilon)\simeq F(k;\epsilon)\sqrt{\dfrac{2\pi}{t|\omega''(k)|}}\cos \left(k\dfrac{x}{\epsilon}+\omega \dfrac{t}{\epsilon}-\dfrac{\pi}{4}\mbox{sign}\,\omega''(k)\right),
\end{equation}
where now $k=k(x,t)$ solves
\begin{equation}
\label{eqk0}
x+\omega'(k)t=0.
\end{equation}
We will now obtain a formula compatible with (\ref{expansion_u_lin})  using the modulation theory.
Let us assume that the amplitude $a$ and the wave number  $k$ are slowly varying functions of space and  time:  
\[
a=a(x,t),\quad  k=k(x,t).\
\]
Plugging the expression
\[
u(x,t;\epsilon)=a(x,t)\cos \left(k(x,t)\dfrac{x}{\epsilon}+\omega(x,t) \dfrac{t}{\epsilon}\right),\quad
\]
into the equation (\ref{linear}) one obtains from the  terms of order one  the equations
\begin{equation}
\label{eq_k_linear}
k_t=\omega'(k)k_x,\quad a_t=\omega'(k)a_x+\dfrac{1}{2}a\omega''(k)k_x,
\end{equation}
which describe the modulation of the wave parameters $a$ and $k$. The curve $\dfrac{dx}{dt}=-\omega'(k)$ is  a characteristic for both the above equations. On such curve
\[
\dfrac{d k}{dt}=0,\quad \dfrac{d a}{dt}=\dfrac{1}{2}a\omega''(k)k_x.
\]
We  look for a self-similar solution of the above equation in  the form
$k=k(z)$ with $z=x/t$. The first equation in (\ref{eq_k_linear}) gives
\[
(z+\omega'(k))k_z=0
\]
which has the solutions  $k_z=0$ or $z+\omega'(k)=0$. This second solution is equivalent to (\ref{eqk0}). Plugging this solution into the equation for the amplitude 
 $a$ one gets
\[
\dfrac{d a}{dt}=-\dfrac{a}{2t},\quad \mbox{or   } a(x,t)=\dfrac{a_0(k)}{\sqrt{t} },
\]
for an arbitrary function $a_0(k)$. Such expression gives an amplitude $a(x,t)$ compatible with the stationary phase asymptotic (\ref{expansion_u_lin}).

\section{Modulation of nonlinear equation}

Now let us consider a similar problem for a nonlinear equation, by adding a nonlinear term $6uu_x$ to the equation (\ref{linear})
\begin{equation}
\label{KdV}
u_t+6uu_x+\epsilon^2 u_{xxx}=0.
\end{equation}
Such equation is called Korteweg de Vries (KdV) equation, and  it describes the behaviour of long waves in shallow water. The coefficient $6$ is front of the nonlinear term, 
is just put for convenience.
The KdV equation admits the travelling wave solution 
\[
u(x,t;\epsilon)=\eta(\phi),\quad \phi=\dfrac{1}{\epsilon}(k x-\omega t+\phi_0),
\]
where we assumed that $\eta$ is a $2\pi$-periodic function of its argument and $\phi_0$ is an arbitrary constant.
Plugging the above ansatz into the KdV equation one obtains after a double integration
\begin{equation}
\label{spectral}
\frac{k^2}{2}\eta_{\phi}^2=-\eta^3+V\eta^2+B\eta+A,\quad V=\dfrac{\omega}{2k}\,,
\end{equation}
where $A$ and $B$ are integration constants and $V$ is the wave velocity.
 In order to get a periodic solution, we assume that the polynomial $-\eta^3+V\eta^2+B\eta+A=-(\eta-e_1)(\eta-e_2)(\eta-e_3)$
with $e_1>e_2>e_3$. Then the periodic motion takes place for $e_2\leq \eta\leq e_1$ and one has the relation
\begin{equation}
\label{spectral1}
k\dfrac{d\eta}{\sqrt{2(e_1-\eta)(\eta-e_2)(\eta-e_3)}}=d\phi,
\end{equation}
so that integrating over a period, one obtains 
\[
2k\int^{e_1}_{e_2}\dfrac{d\eta}{\sqrt{2(e_1-\eta)(\eta-e_2)(\eta-e_3)}}=\oint d\phi=2\pi.
\]
It follows  that the wavenumber $k=\dfrac{2\pi}{L}$ is expressed by  a complete integral of the first kind:
\begin{equation}
\label{k}
k=\pi\dfrac{\sqrt{(e_1-e_3)}}{\sqrt{2}K(m)},\quad m=\dfrac{e_1-e_2}{e_1-e_3},\quad K(m):=\int_0^{\frac{\pi}{2}}\dfrac{d\psi}{\sqrt{1-m^2\sin^2\psi}},
\end{equation}
and  the frequency 
\begin{equation}
\label{omega}
\omega=2k(e_1+e_2+e_3)\,,
\end{equation}
 is obtained by comparison with the polynomial in the r.h.s. of (\ref{spectral}).
Performing an integral between $e_1$ and $\eta$  in equation (\ref{spectral1}) one arrives to the equation
\[
\int_0^{\psi}\dfrac{d\psi'}{\sqrt{1-s^2\sin^2\psi'}}=-\phi\dfrac{\sqrt{e_1-e_3}}{\sqrt{2}k}+K(m),\quad \cos\psi=\dfrac{\sqrt{\eta-e_1}}{\sqrt{e_2-e_1}}.
\]
Introducing  the Jacobi elliptic function   $\mbox{cn}\left(-\phi\dfrac{\sqrt{e_1-e_3}}{\sqrt{2}k}+K(m);m\right)=\cos\psi$ and using  the above equations  we obtain
\begin{equation}
\label{cnoidal}
u(x,t;\epsilon)=\eta(\phi)=e_2+(e_1-e_2)\mbox{cn}^2\left(\dfrac{\sqrt{e_1-e_3}}{\sqrt{2}\epsilon}\left(x-\frac{\omega}{k}t+\dfrac{\phi_0}{k}\right)-K(m);m\right),
\end{equation}
where we use also the evenness of the function $\mbox{cn}(z;m)$.

The function $\mbox{cn}^2(z;m)$ is periodic with period $2K(m)$ and has its maximum at $z=0$ where $\mbox{cn}(0;m)=1$ and its minimum at $z=K(m)$  where $\mbox{cn}(K(m);m)=0$.
Therefore from (\ref{cnoidal}),  the maximum value of the function $u(x,t;\epsilon)$ is  $u_{max}=e_1$ and the minimum value is $u_{\min}=e_2$.
\subsection{Whitham modulation equations}

Now, as we did it in the linear case,  let us suppose  that the integration constants $A$, $B$ and $V$  depend weekly on time and space 
\[
A=A(x,t), \;\;B=B(x,t), \;\;V=V(x,t).
\]
It follows that the wave number and the frequency depends weakly on time  and too.
We are going to derive the equations of $A=A(x,t)$, $B=B(x,t)$ and $V=V(x,t)$ in such a way that  (\ref{cnoidal}) is an approximate  solution of the KdV equation 
(\ref{KdV}) up to sub-leading corrections.  We are going to apply  the nonlinear analogue of the  WKB theory introduced in \cite{DM}.  
For the purpose let us assume that 
\begin{align}
\label{uexp}
u=u(\phi(x,t),x,t),\quad \phi=\dfrac{\theta}{\epsilon}
\end{align}
Pluggin the  ansatz  (\ref{uexp}) into the KdV equation  one has
\begin{equation}
\begin{split}
\label{eq_u}
&u_{\phi}\frac{\theta_t}{\epsilon}+ u_t+6u(u_\phi\frac{\theta_x}{\epsilon}+ u_x)+ \frac{ \theta_x^3}{\epsilon}u_{\phi\phi\phi}+3 \theta_x^2u_{\phi\phi x}+3 \theta_x\epsilon u_{\phi xx }+3\theta_{xx} \epsilon u_{\theta x}\\
&+3 \theta _{xx}\theta_x u_{\phi \phi}+\theta_{xxx} \epsilon u_\phi+\epsilon^2u_{xxx}=0.
\end{split}
\end{equation}
Next assuming that  $u$ has an expansion in power of $\epsilon$, namely $u=u_0+\epsilon u_1+\epsilon^2u_2+\dots$ one obtain from (\ref{eq_u})  at order $1/\epsilon$

%
\[
\theta_t u_{0,\phi}+6\theta_x u_0u_{0,\phi}+\theta_x^3u_{0,\phi\phi\phi}=0.
\]
The above equation  gives the cnoidal wave solution (\ref{cnoidal}) if $u_0(\phi)=\eta(\phi)$ and 
\begin{equation}
\label{phiXT}
\theta_t=-\omega,\quad \theta_x=k,
\end{equation}
where  $k$  and $\omega$ are the frequency and wave number of the cnoidal wave as defined in (\ref{k}) and (\ref{omega}) respectively. 
Compatibility of equation (\ref{phiXT}) gives
\begin{equation}
\label{whitham00}
k_t+\omega_x=0,
\end{equation}
which is the first equation we are looking for.
To obtain the other equations  let us introduce the linear operator
\[
\mathcal{L}:=\omega \dfrac{\partial}{\partial \phi}-6k\dfrac{\partial}{\partial \phi}u_0-k^3\dfrac{\partial^3}{\partial \phi^3},
\]
with formal adjoint $\mathcal{L}^{\dagger}=\omega \dfrac{\partial}{\partial \phi}-6ku_0\dfrac{\partial}{\partial \phi}-k^3\dfrac{\partial^3}{\partial \phi^3}.$
Then at order $\epsilon^0$  equation (\ref{eq_u}) gives 
\begin{align*}
\mathcal{L}u_1=&R(u_0),\quad R(u_0):=u_{0,t}+6u_0 u_{0,x}+3 \theta_x^2u_{0,\phi\phi x}+3 \theta_{xx}\theta_x u_{0,\phi \phi}.
\end{align*}
In a similar way it is possible to 
get the equations for the higher order correction terms.
A condition of solvability of the above  equation can be obtained by observing that the  integral over a period     of the l.h.s of the above equation against  the constant function and the function $u_0$  is equal to zero 
because $1$ and $u_0$ are in the kernel of $\mathcal{L}^{\dagger}$.
Therefore it follows that 
\[
0=\int_0^{2\pi}R(u_0)d\phi=\partial_t\int_0^{2\pi}u_0d\phi+3\partial_x\int_0^{2\pi}u^2_0d\phi
\]
and 
\begin{align*}
0=\int_0^{2\pi}u_0R(u_0)d\phi=&\partial_t\int_0^{2\pi}\dfrac{1}{2}u^2_0d\phi+2\partial_x\int_0^{2\pi}u^3_0d\phi\\
&+3 \int_0^{2\pi} u_0(\theta_x^2u_{0,\phi\phi x}+ \theta_{xx}\theta_x u_{0,\phi \phi})d\phi.
\end{align*}
By denoting with the bracket $\langle\,.\,\rangle$ the average over a period,
we rewrite  the above two equations, after elementary algebra and an integration by parts,  in the form 
\begin{align}
\label{whitham1}
&\partial_t\langle u_0\rangle +3\partial_x\langle  u^2_0\rangle=0 \\
\label{whitham2}
&\partial_t\langle u^2_0\rangle+4\partial_x\langle u^3_0\rangle-3\partial_x\langle \theta_x^2 u_{0,\phi}^2\rangle=0.
\end{align}
   Using the identities
   \[
 \langle u_0 u_{0,\phi\phi}+u_{0,\phi}^2\rangle =0,\quad \langle u_{0,\phi\phi} \rangle =0,
  \] 
  and (\ref{spectral}), we obtained the identities for the elliptic integrals
  \[
  \int_{e_1}^{e_2}\dfrac{5\eta^3-4V\eta^2-3B\eta-2A}{\sqrt{-\eta^3+V\eta^2+B\eta+A}}d\eta=0,\quad  \int_{e_1}^{e_2}\dfrac{-3\eta^2+2V\eta+B}{\sqrt{-\eta^3+V\eta^2+B\eta+A}}d\eta=0.
    \]
    Introducing the integral $W:=\frac{\sqrt{2}}{\pi}\int_{e_1}^{e_2}\sqrt{-\eta^3+V\eta^2+B\eta+A}d\eta$ and 
    using  the above two identities and  the relations
  $  kW_A=1$, $\langle u_0\rangle= 2\pi kW_B$ and $\langle u_0^2\rangle= 2\pi kW_V$   where $W_A$, $W_B$  and $W_V$ are the partial derivatives of $W$ with respect to $A$, $B$ and $V$ respectively,
     we can reduce  (\ref{whitham00}),  (\ref{whitham1}) and (\ref{whitham2})  to the form
   \begin{align}
   \label{whitham000}
&    \dfrac{\partial}{\partial t}W_A+2V\dfrac{\partial}{\partial x}W_A-2W_A\dfrac{\partial}{\partial x}V=0\\
      \label{whitham11}
   & \dfrac{\partial}{\partial t}W_B+2V\dfrac{\partial }{\partial x}W_B+W_A\dfrac{\partial }{\partial x}B=0\\
\label{whitham22}
  &   \dfrac{\partial}{\partial t}W_V+2V\dfrac{\partial }{\partial x}W_V-2W_A\dfrac{\partial }{\partial x}A=0.
  \end{align}
 The equation (\ref{whitham000}), (\ref{whitham11}) and (\ref{whitham22})   are   the Whitham modulation equations for the  parameters $A$, $B$ and $V$.
   The same equations can  also be derived according to Whitham's original ideas  of  averaging method  applied to conservation laws, to Lagrangian or to Hamiltonians \cite{W}. 
Using $e_1$, $e_2$ and $e_3$ as independent variables, instead of their symmetric function $A$, $B$ and $V$, Whitham  reduced the above three equations to the form

%
\begin{equation}
\label{whitham0}
\dfrac{\partial}{\partial t}e_i+\sum_{k=1}^3\sigma_i^k\dfrac{\partial}{\partial x}e_k=0,\quad i=1,2,3,
\end{equation}
for the  matrix $\sigma_i^k$ given by 
\[
\sigma=2V-W_A\begin{pmatrix}
\partial_{e_1}W_A&\partial_{e_2}W_A&\partial_{e_3}W_A\\
\partial_{e_1}W_B&\partial_{e_2}W_B&\partial_{e_3}W_B\\
\partial_{e_1}W_V&\partial_{e_2}W_B&\partial_{e_3}W_V
\end{pmatrix}^{-1}
\begin{pmatrix}
2&2&2\\
e_2+e_3&e_1+e_3&e_1+e_2\\
2e_2e_3&2e_1e_3&2e_1e_2\end{pmatrix},
\]
where $\partial_{e_i}W_A$ is the partial derivative with respect to $e_i$ and the same notation holds for the other quantities.
Equations (\ref{whitham0}) is a system of quasi-linear equations for $e_i=e_i(x,t)$, $j=1,2,3$. Generically, a quasi-linear  $3\times 3$ system cannot be reduced to a  diagonal form. However  Whitham, analyzing the form of the matrix $\sigma$,   was able to get the Riemann invariants that reduce the system to diagonal form. Indeed making the change of coordinates  
\begin{equation}
\label{Riemann}
\beta_1=\dfrac{e_2+e_1}{2},\;\;\beta_2=\dfrac{e_1+e_3}{2},\;\;\beta_3=\dfrac{e_2+e_3}{2},
\end{equation}
with 
\[
\beta_3<\beta_2<\beta_1,
\]
 the Whitham modulation equations (\ref{whitham0}) are diagonal and take the form
\begin{equation}
\label{Whitham}
\dfrac{\partial}{\partial t}\beta_i+\lambda_i \dfrac{\partial}{\partial x}\beta_i=0,\quad i=1,2,3,
\end{equation}
where the characteristics speeds  $\lambda_i=\lambda_i(\beta_1,\beta_2,\beta_3)$ are
\begin{align}
\label{speed}
&\lambda_i=2(\beta_1+\beta_2+\beta_3)+4\dfrac{\prod_{i\neq k}(\beta_i-\beta_k)}{\beta_i+\alpha},\\
\label{alpha}
&\alpha=-\beta_1+(\beta_1-\beta_3)\dfrac{E(m)}{K(m)},\quad m=\dfrac{\beta_2-\beta_3}{\beta_1-\beta_3},
\end{align}
where $E(m)=\int_0^{\pi/2}\sqrt{1-m\sin \psi^2}d\psi$ is the complete elliptic integral of the second kind.
Another compact form of the Whitham modulations equations (\ref{Whitham}) is 
\begin{equation}
\label{Whithamcompact}
\dfrac{\partial k}{\partial \beta_i}\dfrac{\partial \beta_i}{\partial t}+\dfrac{\partial \omega}{\partial \beta_i}\dfrac{\partial \beta_i}{\partial x}=0,\quad i=1,2,3,
\end{equation}
where the above equations do  not contain the  sum over repeated indices. Observe that the above expression can be derived from the conservation of waves (\ref{whitham00}) by assuming that the Riemann invariants $\beta_1>\beta_2>\beta_3$  vary independently.
Such form  (\ref{Whithamcompact}) is quite  general and easily adapts to other modulation  equations  ( see for example the book \cite{Kamchatnov}).
The equations (\ref{Whithamcompact}) gives another  expression for the speed $\lambda_i=2(\beta_1+\beta_2+\beta_3)+2\dfrac{k}{\partial_{\beta_i}k}$ which was obtained in \cite{GKE}.

The Whitham equations are a systems of $3\times 3$  quasi-linear hyperbolic equations  namely for $\beta_1>\beta_2>\beta_3$ one has \cite{L}
\[
\lambda_1>\lambda_2>\lambda_3.
\]
Using the expansion of the elliptic integrals as $m\to 0$   (see e.g. \cite{Lawden})
\begin{equation}
\label{elliptictrail}
K(m)=\dfrac{\pi}{2}\left(1+\dfrac{m}{4}+\dfrac{9}{64}m^2+O(m^3)\right),\quad
E(m)=\dfrac{\pi}{2}\left(1-\dfrac{m}{4}-\dfrac{3}{64}m^2+O(m^3)\right),
\end{equation}
and $m\to 1$
\begin{equation}
\label{ellipticlead}
E(m)\simeq 1+\dfrac{1}{2}(1-\sqrt{m})\left[\log\dfrac{16}{1-m}-1\right],
\quad K(m)\simeq \dfrac{1}{2}\log\dfrac{16}{1-m},
\end{equation}
 one can verify that the speeds $\lambda_i$ have the following limiting behaviour respectively
\begin{itemize}
\item at $\beta_2=\beta_1$
\begin{equation}
\begin{split}
\label{limit_edge}
&\lambda_1(\beta_1,\beta_1,\beta_3)=\lambda_2(\beta_1,\beta_1,\beta_3)=4\beta_1+2\beta_3\\
&\lambda_3(\beta_1,\beta_1,\beta_3)=6\beta_3;
\end{split}
\end{equation}
\item at $\beta_2=\beta_3$ one has
\begin{equation}
\begin{split}
\label{limit_trail}
&\lambda_1(\beta_1,\beta_3,\beta_3)=6\beta_1\\
&\lambda_2(\beta_1,\beta_3,\beta_3)=\lambda_3(\beta_1,\beta_3,\beta_3)=12\beta_3-6\beta_1.
\end{split}
\end{equation}
\end{itemize}
Namely, when $\beta_1=\beta_2$, the equation for 
$\beta_3$ reduces to the Hopf equation $\dfrac{\partial }{\partial t}\beta_3 +6\beta_3\dfrac{\partial }{\partial x} \beta_3=0$.
In the same way when $\beta_2=\beta_3$ the equation for $\beta_1$ reduces to the Hopf equation.

In the coordinates $\beta_i$,  $i=1,2,3$ the travelling wave solution  (\ref{cnoidal}) takes the form
 \begin{equation}
\label{cnoidal2}
u(x,t;\epsilon)=\beta_1+\beta_3-\beta_2+2(\beta_2-\beta_3)\mbox{cn}^2\left(K(m)\dfrac{\Omega}{\pi \epsilon} +K(m);m\right),
\end{equation}
where
\begin{equation}
\label{Omega}
\Omega:=kx-\omega t+\phi_0=\pi\dfrac{\sqrt{\beta_1-\beta_3}}{K(m)}(x-2t(\beta_1+\beta_2+\beta_3))+\phi_0,\quad m=\dfrac{\beta_2-\beta_3}{\beta_1-\beta_3}.
\end{equation}
We recall that 
\begin{equation}
\label{kappa_omega}
k=\pi \dfrac{\sqrt{\beta_1-\beta_3}}{K(m)},\quad \omega=2k(\beta_1+\beta_2+\beta_3),
\end{equation}
are the wave-number and frequency of the oscillations respectively.

In the formal  limit $\beta_1\to \beta_2$, the above cnoidal wave reduce to the soliton solution since $\mbox{cn}(z,m)\overset{m\to 1}{\to} {\rm sech}(z)$, while  the limit $\beta_2\to\beta_3$ is the small amplitude limit where the oscillations become linear and $\mbox{cn}(z,m)\overset{m\to 0}{\to} \cos (z)$. Using identities among elliptic functions \cite{Lawden}  we can  rewrite the travelling wave solution (\ref{cnoidal2}) using theta-functions  
 \begin{equation}
\label{elliptic0}
u(x,t,\epsilon)=\beta_1+\beta_2+\beta_3+2\alpha+ 2\epsilon^2\frac{\partial^2}{\partial
x^2}\log\vartheta\left(\dfrac{\Omega(x,t)}{2 \pi \epsilon};\tau\right),
\end{equation}
 with $\alpha$ as in (\ref{alpha}) and where for any $z\in\mathbb{C}$ the function
 $\vartheta(z;\tau)$ is defined by the
Fourier series
\begin{equation}
\label{tau}
\vartheta(z;\tau)=\sum_{n\in\mathbb{Z}}e^{\pi i n^2\tau+2\pi i nz},\quad \tau= i\dfrac{K'(m)}{K(m)}.
\end{equation}
The formula (\ref{elliptic0})  is a particular case of the Its-Matveev formula \cite{ItsMatveev} that describes  the quasi-periodic solutions of the KdV equation through higher order $\theta$-functions.\\

{\bf Remark 2.1}
We remark that for fixed $\beta_1,\beta_2$ and $\beta_3$, formulas (\ref{cnoidal2}) or (\ref{elliptic0}) give an exact solution of the KdV equation (\ref{KdV}), while when 
$\beta_j=\beta_j(x,t)$ evolves according to the Whitham equations, such formulas give an approximate solution of the KdV equation (\ref{KdV}).
We also remark that  in the derivation of the Whitham equations, we did not get any information for an eventual modulation of the arbitrary phase $\phi_0$.
The modulation of the phase requires a higher order analysis, that won't be explained here. However we will give below a formula for the phase.
\\

{\bf Remark 2.2}
The Riemann invariants $\beta_1$, $\beta_2$ and $\beta_3$ have an important spectral meaning.
Let us  consider the spectrum of the Schr\"odinger equation
\[
\epsilon^2\dfrac{d^2}{dx^2}\Psi+u\Psi=-\lambda\Psi,
\]
where $u(x,t;\epsilon)$ is a solution of the KdV equation. The main discovery of Gardener, Green Kruskal and Miura \cite{GGKM} is that the spectrum of the Schr\"odinger operator is constant in time 
if  $u(x,t;\epsilon)$  evolve according to the KdV equation. This  important observation is the starting point of inverse scattering and  the modern theory of integrable systems in infinite dimensions.

If $u(x,t;\epsilon)$ is the travelling wave solution (\ref{elliptic0}), where $\beta_1>\beta_2>\beta_3$ are constants, then the  Schr\"odinger equation coincides with the Lam\'e equation 
and its  spectrum coincides with the Riemann invariants $\beta_1>\beta_2>\beta_3$. The stability zones of the spectrum are the bands $(-\infty,\beta_3]\cup[\beta_2,\beta_1]$.
The corresponding solution $\Psi(x, t;\lambda)$  of the Schr\"odinger equation is quasi-periodic in $x$ and $t$ with monodromy
\[
\Psi(x+\epsilon L, t;\lambda)=e^{ip(\lambda) L}\Psi(x,t;\lambda)
\] 
and 
\[
\Psi(x, t+\epsilon T;\lambda)=e^{iq(\lambda) T}\Psi(x,t;\lambda),
\]
where $\epsilon L$ and $\epsilon T$ are the wave-length and the  period of the oscillations.
The functions $p(\lambda)$ and $q(\lambda)$ are called quasi-momentum and quasi-energy and for the cnoidal wave solution they take the simple form
\[
p(\lambda)=\int_{\beta_2}^\lambda dp(\lambda'),\quad q(\lambda)=\int_{\beta_2}^\lambda dq(\lambda'),
\]
where $dp$ and $dq$ are given by the expression
\[
dp(\lambda)=\dfrac{(\lambda+\alpha)d\lambda}{2\sqrt{(\beta_1-\lambda)(\lambda-\beta_2)(\lambda-\beta_3)}},\quad dq(\lambda)=12\dfrac{(\lambda^2-\frac{1}{2}(\beta_1+\beta_2+\beta_3)\lambda+\gamma)d\lambda}{2\sqrt{(\beta_1-\lambda)(\lambda-\beta_2)(\lambda-\beta_3)}}
\]
with  the constant $\alpha$ defined in (\ref{alpha}) and  $\gamma=\dfrac{\alpha}{6}(\beta_1+\beta_2+\beta_3)+\dfrac{1}{3}(\beta_1\beta_2+\beta_1\beta_3+\beta_2\beta_3)$
Note that the constants  $\alpha$ and $\gamma$ are chosen so that 
\[
\int_{\beta_3}^{\beta_2}dp=0,\quad  \int_{\beta_3}^{\beta_2}dq=0.
\]
The square root $\sqrt{(\beta_1-\lambda)(\lambda-\beta_2)(\lambda-\beta_3)}$ is analytic in the complex place $\mathbb{C}\backslash\{ (-\infty,\beta_3]\cup[\beta_2,\beta_1]\}$ and real for large negative $\lambda$ so that $p(\lambda)$ and $q(\lambda)$ are real in the stability zones.
The Whitham modulation equations (\ref{Whitham}) are equivalent to 
\begin{equation}
\label{diffWhitham}
\dfrac{\partial }{\partial t}dp(\lambda)+\dfrac{\partial}{\partial x}dq(\lambda)=0,
\end{equation}
for any $\lambda$.
Indeed by multiplying  the above equation by $(\lambda-\beta_i)^{\frac{3}{2}}$ and taking  the limit $\lambda\to \beta_i$, one gets (\ref{Whitham}).
Furthermore
\[
k=\int_{\beta_2}^{\beta_1}dp,\quad \omega=\int_{\beta_2}^{\beta_1}dq,
\]
 with $k$ and $\omega$ the wave-number and frequency as in (\ref{kappa_omega}), so that integrating (\ref{diffWhitham})  between $\beta_1$ and $\beta_2$ and observing that the integral does not depend on the path of integration one recovers the equation of wave conservation  (\ref{whitham00}).

\section{Application of Whitham modulation equations }
As in the linear case, the modulation equations have important applications in the description of the solution of the Cauchy problem of the  KdV equation in asymptotic limits.
Let us consider the initial value problem
\begin{equation}
\label{KdV1}
\left\{\begin{array}{ll}
&u_t+6uu_x+\epsilon^2 u_{xxx}=0\\
&u(x,0;\epsilon)=f(x),
\end{array}\right.
\end{equation}
where $f(x)$ is an initial data independent from $\epsilon$.
When we study  the solution of such initial value problem $u(x,t;\epsilon)$ one can  consider  two limits:
\begin{itemize}
\item the long time behaviour, namely
\[
u(x,t;\epsilon)\overset{t\to \infty}{\to}\;?,\quad 
\mbox{$\epsilon$ fixed};
\]
\item the small dispersion limit, namely 
\[
u(x,t;\epsilon)\overset{\epsilon\to 0}{\to}\;?,\quad 
\mbox{$x$ and $t$ in compact sets.}
\]
\end{itemize}
These two limits have been widely studied in the literature.
The physicists Gurevich and Pitaevski \cite{GP1}  were among the first to address these limits and      gave an heuristic solution imitating  the linear case.
Let us first consider  one of the case studied  by Gurevich and Pitaevski, namely  a decreasing step initial data
\begin{equation}
\label{step}
f(x)=\left\{\begin{array}{ll}
c&\;\; \mbox{for}\;\; \;x<0,\;\;c>0,\\
0& \;\;\mbox{for}\;\;\;x>0.
\end{array}\right.
\end{equation}
Using the Galileian  invariance of KdV equation, namely $x\to x+6Ct$, $t\to t$ and $u\to u+C$, every  initial data with a single step can be reduced to the above  form.
The above  step  initial data is invariant under the rescaling $x/\epsilon\to x$ and $t/\epsilon\to$, therefore, in this particular case it is completely equivalent to study the small $\epsilon$ asymptotic,  or the long time asymptotics of the solution.

Such initial data is called compressive step, and the  solution of the Hopf equation  $v_t+6vv_x=0$  ($\epsilon=0$ in (\ref{KdV1}) )  develop a shock for $t>0$.
The shock front  $s(t)$ moves with velocity $3ct$ while the 
multi-valued  piece-wise continuos solution of the Hopf equation $v_t+6vv_x=0$  for the same initial data  is given by
\[
v(x,t)=\left\{
\begin{array}{ll}
c&\;\;\mbox{for}\;x<6tc,\\
&\\
\dfrac{x}{6t}&\;\; \mbox{for} \;0\leq x\leq 6tc,\\
&\\
0&\;\;\mbox{for}\;x\geq 0.
\end{array}\right.
\]

For $t>0$ the solution $u(x,t;\epsilon)$  of the KdV equation develops a train of oscillations near the discontinuity.
 These oscillations   are  approximately described by the travelling wave solution (\ref{elliptic0})  of the KdV equation where  $\beta_i=\beta_i(x,t)$, $i=1,2,3$,  evolve according to the Whitham equations.
However one needs to fix the solution of the Whitham equations.  Given the self-similar structure of the solution of the Hopf equation, it is natural to 
look for a self-similar solution of the Whitham equation in the form $\beta_i=\beta_i(z)$ with $z=\dfrac{x}{t}$.
Applying this change of variables to the Whitham equations one obtains
\begin{equation}
\label{Whithamself}
(\lambda_i-z)\dfrac{\partial \beta_i}{\partial z}=0,\quad i=1,2,3,
\end{equation}
whose solution is $\lambda_i=z$ or  $\partial_z\beta_i=0$.
A natural request that follows from the relations  (\ref{limit_edge})  and  (\ref{limit_trail})  is that at   the  right boundary of the oscillatory zone  $z_+$, when   
$\beta_1(z_+)=\beta_2(z_+)$, the function $\beta_3$ has to match the Hopf solution that is constant and equal to zero, namely $\beta_3(z_+)=0$.
Similarly, at the left boundary $z_-$ when $\beta_2(z_-)=\beta_3(z_-)$,  the function $\beta_1(z_-)=c$ so that it matches the Hopf solution. 
From these observations  it follows that the solution of (\ref{Whithamself})  for $z_-\leq z\leq z_+$ is given by
\begin{equation}
\label{solself}
\beta_1(z)=c,\quad \beta_3(z)=0,\quad z=\lambda_2(c,\beta_2,0).
\end{equation}
 In order to determine the values $z_{\pm}$ it is sufficient to let $\beta_2\to c$ and $\beta_2\to 0$ respectively  in the  last equation in (\ref{solself}).
Using the relations  (\ref{limit_edge}) and (\ref{limit_trail}) one  has $\lambda_2(c,c,0)=4c$ and $\lambda_2(c,0,0)=-6c$ so that 
\[
z_-=-6c,\;\;\mbox{or} \;\;x_-(t)=-6ct \quad \mbox{and   }\quad z_+=4c,\;\;\mbox{or}\;\;\;x_+(t)=4ct.
\]
According to Gurevich and Pitaevski   for $-6ct<x<4t$ and $t\gg 1$, the asymptotic solution of the Korteweg de Vries equation with step initial data (\ref{step})  is given by the modulated travelling wave solution (\ref{cnoidal2}), 
namely

\begin{equation}
\label{longtime}
u(x,t;\epsilon)\simeq c-\beta_2+2\beta_2\,\mbox{cn}^2\left(\dfrac{\sqrt{c}}{\epsilon}(x-2t(c+\beta_2)) +\dfrac{K(m)}{\pi\epsilon}\phi_0 +K(m);m\right),
\end{equation}
with 
\[
 m=\dfrac{\beta_2(x,t)}{c},
 \]
 where $\beta_2(x,t)$ is given by (\ref{solself}). The phase $\phi_0$ in (\ref{longtime}) has not been described by Gurevich and Pitaevski.
Finally in the remaining regions of the $(x,t>0)$ one has 
\[
u(x,t,\epsilon)\simeq\left\{\begin{array}{ll}
c&\mbox{for}\;\;x<-6ct,\\
0&\mbox{for}\;\;x>4ct.
\end{array}\right.
\]
This  heuristic description has been later  proved  in a rigorous mathematical way (see the next section).
We remark that at the  right boundary  $x_+(t)$ of the oscillatory zone, when $\beta_2\to c$, $\beta_1\to c$ and  $\beta_3\to 0$, the cnoidal wave (\ref{longtime})  tends to a soliton,  
$cn(z;m)\to {\rm sech} z$ as $m\to 1$.
\begin{figure}
  \includegraphics[width=1.0\textwidth]{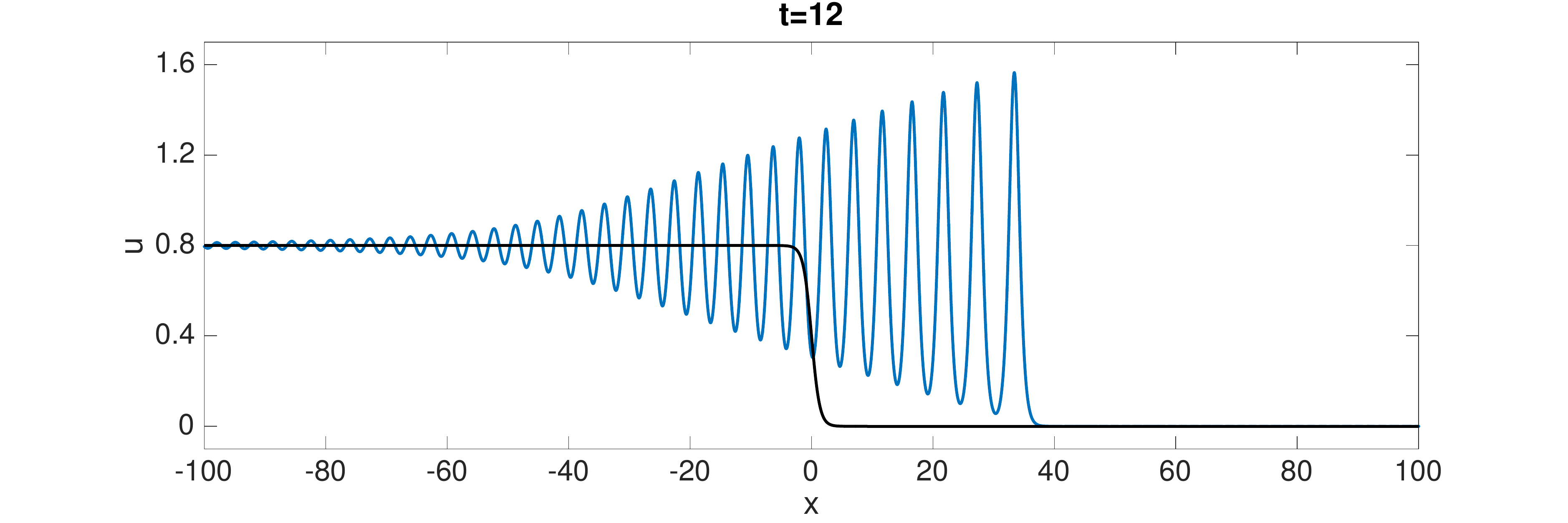}    
  \caption{ In black the initial data  (a smooth step) and in blue KdV  solution at time $t=12$ and $\epsilon=1$.  One can clearly see the  height of the rightmost  oscillation (approximately a soliton) is about two times the height of the initial  step}
 \label{figstep}
\end{figure}

\begin{figure}[H]
  \includegraphics[width=0.9\textwidth]{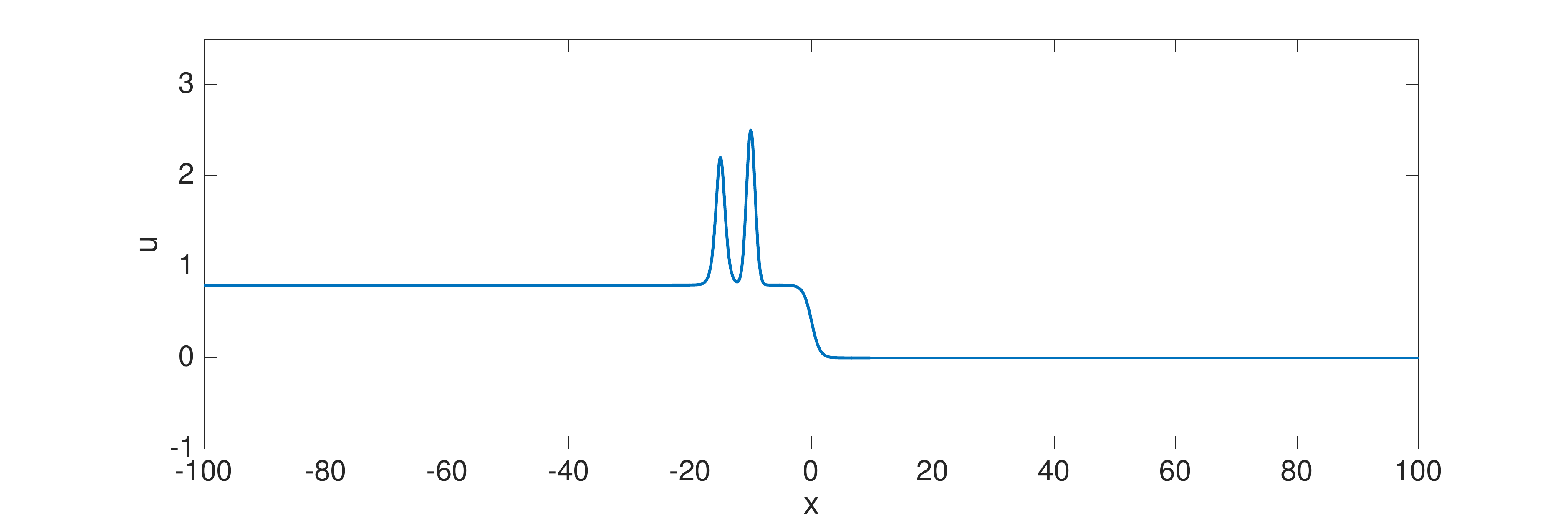}    \includegraphics[width=0.9\textwidth]{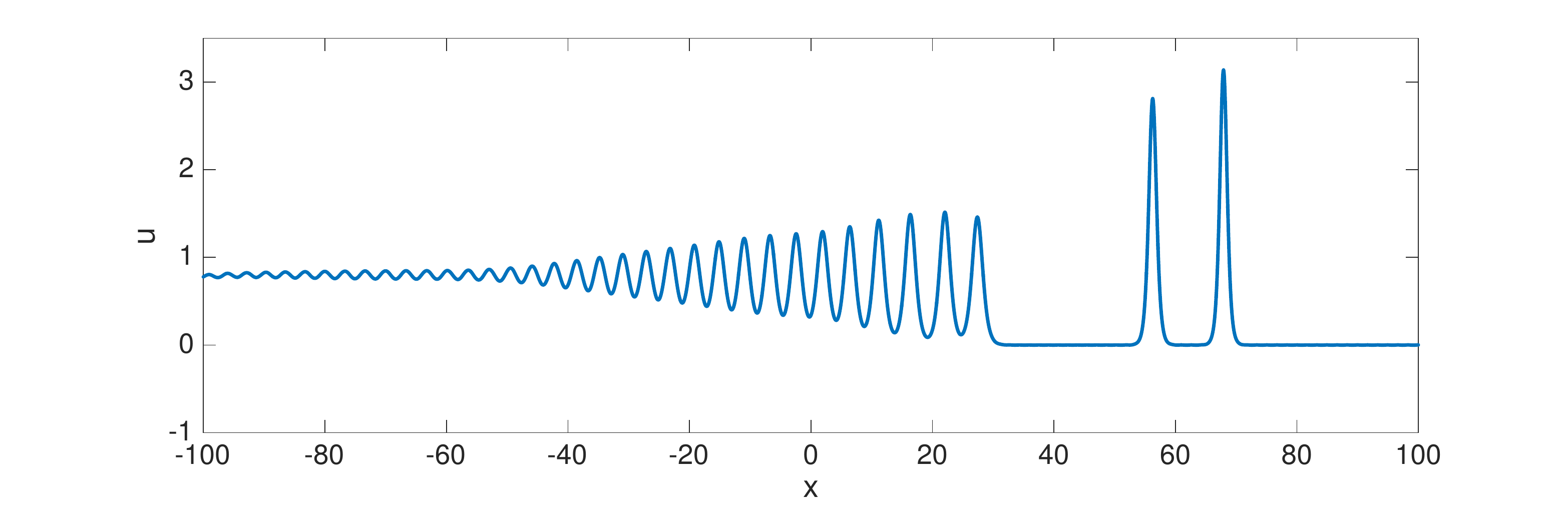}  
  \caption{ On top  the  step-like  initial data and on bottom the solution at time $t=12$.  One can clearly see  the soliton region containing two solitons  and the collision-less  shock region where modulated oscillations are formed.}
 \label{figl}
\end{figure}

Using the relation $x_+(t)=4ct$, the limit of the elliptic solution (\ref{longtime})  when $\beta_2\to\beta_1\to c$ gives
\begin{equation}\label{expansion formal}
u(x,t,\epsilon)\simeq 2c\,\,{\rm sech}^2\left[  \dfrac{x-x_+(t)}{\epsilon}\sqrt{c}+\dfrac{1}{2}\log\left(\dfrac{16c}{c-\beta_2}\right)+\dfrac{\tilde{\phi}_0}{\epsilon}\right],
\end{equation}
where the logarithmic term is due to the expansion  of the complete elliptic integral $K(m)$ as in (\ref{ellipticlead}) and $c-\beta_2=O(\epsilon)$.
The  determination of the limiting value of the  phase $\tilde{\phi}_0$   requires a deeper analysis \cite{CG3}.
The important feature of the above formula is that if the argument of the $\rm{sech}$ term is approximately zero near the point $x_+(t)$, then   the height of the rightmost oscillation is twice the initial step $c$.
This occurs for a single step initial data (see figure~\ref{figstep}) while for  step-like initial data as  in figure~\ref{figl} this is  clearly less evident.

The Gurevich Pitaevsky problem has been  studied also for  perturbations of the KdV equation with forcing, dissipative or conservative non integrable terms \cite{El},\cite{Kamchatnov},\cite{Kamchatnov1}  and applied to the evolution  of solitary waves and undular bores
in shallow-water flows over a gradual slope with bottom friction \cite{El2}.  

\subsection{Long time asymptotics}
The study of the  long time asymptotic of the KdV solution   was initiated around 1973 with the work of Gurevich and Pitaevski \cite{GP1} for step-initial data and Ablowitz and Newell \cite{AblowitzNewell} for rapidly decreasing initial data.
 By that time it was clear  that for rapidly decreasing initial data 
the solution of the  KdV   equation splits into a number of solitons moving to the right and a decaying radiation moving to the left. The first numerical evidence of such behaviour was found by Zabusky and Kruskal \cite{ZabuskyKruskal}.
The first mathematical results were given by Ablowitz and Newell \cite{AblowitzNewell} and  Tanaka \cite{Tanaka} for rapidly decreasing initial data.  Precise asymptotics  on the radiation part were first   obtained by Zakharov and Manakov, \cite{ZM}, Ablowitz and Segur \cite{AblowitzSegur} and Buslaev and Sukhanov \cite{BS}, Venakides \cite{V3}. Rigorous mathematical results were also  obtained by  Deift and Zhou \cite{DZ}, inspired by earlier work  by Its \cite{Its}; see also the review \cite{DIZ}  and the book \cite{Schuur} for the history of the problem.
In \cite{AblowitzSegur}, \cite{GP2}   the  region  with modulated oscillations of order O(1) emerging in the long time asymptotics    was called {\it collisionless shock region}. 
In the physics and applied mathematics literature such oscillations are also called dispersive shock waves, dissipationless shock wave or  undular bore.
The phase of the oscillations was obtained in \cite{DVZ1}.
 Soon after the Gurevich and Pitaevski's paper,  Khruslov  \cite{Khruslov} studied the long time asymptotic of  KdV via inverse scattering for step-like initial data.
In more recent works, using the techniques introduced in \cite{DZ}, the long time asymptotic of KdV solution has been obtained for  step like initial data improving some  error estimates obtained earlier and  with the determination of the phase $\phi_0$ of the oscillations \cite{Teschl1}, see also \cite{AB}.  Long time asymptotic of KdV with different boundary conditions at infinity has been considered in \cite{Bikbaev}. The long time asymptotic of the expansive step has been considered in \cite{LN}.

Here we report  from \cite{Teschl1}  about the  long time asymptotics of KdV  with step like initial data  $f(x)$, namely initial data converging rapidly to the limits
\begin{equation}
\label{asym}
\left\{
\begin{array}{ll}
f(x)\to 0&\mbox{for}\;\;x\to +\infty\\
f(x)\to c>0&\mbox{for}\;\;x\to -\infty,
\end{array}\right.
\end{equation}
but in the finite region of the $x$ plane any kind of regular  behaviour is allowed. The initial data has to satisfy the extra technical assumption of being sufficiently smooth.
Then the  asymptotic behaviour of $u(x,t;\epsilon)$ for fixed $\epsilon$ and $t\to \infty$ has been obtained applying the Deift-Zhou method in  \cite{DZ}:
\begin{itemize}
\item  in the region  $x/t>4c+\delta$, for some $\delta>0$,  the solution is asymptotically given by the sum of solitons if the initial data contains solitons otherwise the solution is approximated by zero at leading order;
\item in the region  $-6c+\delta_1<x/t<4c-\delta_2$, for some $\delta_1,\delta_2>0$,  ({\it collision-less shock region}) the solution $u(x,t;\epsilon)$  is given by the modulated travelling wave  (\ref{longtime}), or using $\vartheta$-function by 
 (\ref{elliptic0}), namely 
\begin{equation}
\label{longtime1}
u(x,t;\epsilon)= \beta_2(x,t)-c+2c\dfrac{E(m)}{K(m)}+\dfrac{2k^2}{(2\pi)^2}\left(\log\vartheta\left(\dfrac{kx-\omega t+\phi_0}{2\pi \epsilon};\tau\right)\right)''+o(1)
\end{equation}
where
\[
k=\pi \dfrac{\sqrt{c}}{ K(m)}, \quad \omega=2k(c+\beta_2),\quad m=\dfrac{\beta_2(x,t)}{c}
\]
with $\beta_2=\beta_2(x,t)$ determined by (\ref{solself}).  In the above formula  the prime in the $\log\vartheta$ means derivative  with respect to the argument, namely 
 $(\log\vartheta(z_0;\tau))''=\dfrac{d^2}{dz^2}\log\vartheta(z+z_0;\tau))|_{z=0}$.
The phase $\phi_0$ is 
\begin{equation}
\label{phaseL}
\phi_0=\dfrac{k}{\pi}\int_{\beta_2}^c\dfrac{\log|\bar{T}(i\sqrt{z})T_1(i\sqrt{z})|dz}{\sqrt{z(c-z)(z-\beta_2)}},
\end{equation}
where $T$ and $T_1$ are the transmission coefficients  of the Schr\"odinger equation $\epsilon^2\dfrac{d^2}{dx^2}\Psi+f(x)\Psi=-\lambda\Psi$ from the right and left respectively.

{\it The  remarkable feature of formula (\ref{longtime1})  is that the   description of the  collision-less shock region   for step-like initial data coincides with the formula obtained  by Gurevich and Pitaevsky 
for  the single step initial data  (\ref{step})  up to a  phase factor. Indeed   the initial data is entering explicitly through the transmission coefficients only in  the phase $\phi_0$ of the oscillations.}
\item In the region $x/t<-6t-\delta_3$, for some constant $\delta_3>0$,  the solution is asymptotically close to  the background $c$ up to a decaying linear oscillatory term.
\end{itemize}

We remark that  the higher order correction  terms of the  KdV solution in the large $t$ limit  can be found in \cite{AblowitzSegur}, \cite{BS}, \cite{Teschl1},  \cite{ZM}. 
For example in  the region $x<-6tc$ the solution is asymptotically close to the background $c$ up to a decaying linear oscillatory term.
We also  remark that the boundaries of the above three regions of the $(x,t)$ plane have escaped our  analysis.
In such regions the asymptotic description of the KdV solution  is given by elementary functions or Painlev\'e trascendents  see \cite{SA} or the more recent work \cite{BIS}.

The technique introduced by Deift-Zhou  \cite{DZ} to study asymptotics for integrable equations has proved to be very powerful and effective to study  asymptotic behaviour  of many other integrable equations
like for example the semiclassical limit of the focusing nonlinear Schr\"odinger equation  \cite{KMM}, the long time asymptotics of the Camassa-Holm equation \cite{BIS}  or the long time asymptotic of the perturbed defocusing nonlinear Schr\"odinger equation \cite{DZ2}.

\subsection{Small $\epsilon$ asymptotic}
The idea of the formation of an oscillatory structure in the limit of small dispersion of a  dispersive equation belongs to Sagdeev \cite{Sagdeev}.
Gurevish and Pitaevskii in 1973  called the oscillations, arising in the small dispersion limit of KdV, {\it  dispersive shock waves} 
 in analogy with the shock waves appearing in the zero dissipation limit of the Burgers  equation.
A very recent  experiment in a water tank  has been set up where the dispersive shock waves have been reproduced \cite{OT}.

\begin{figure}
   \includegraphics[width=0.9\textwidth]{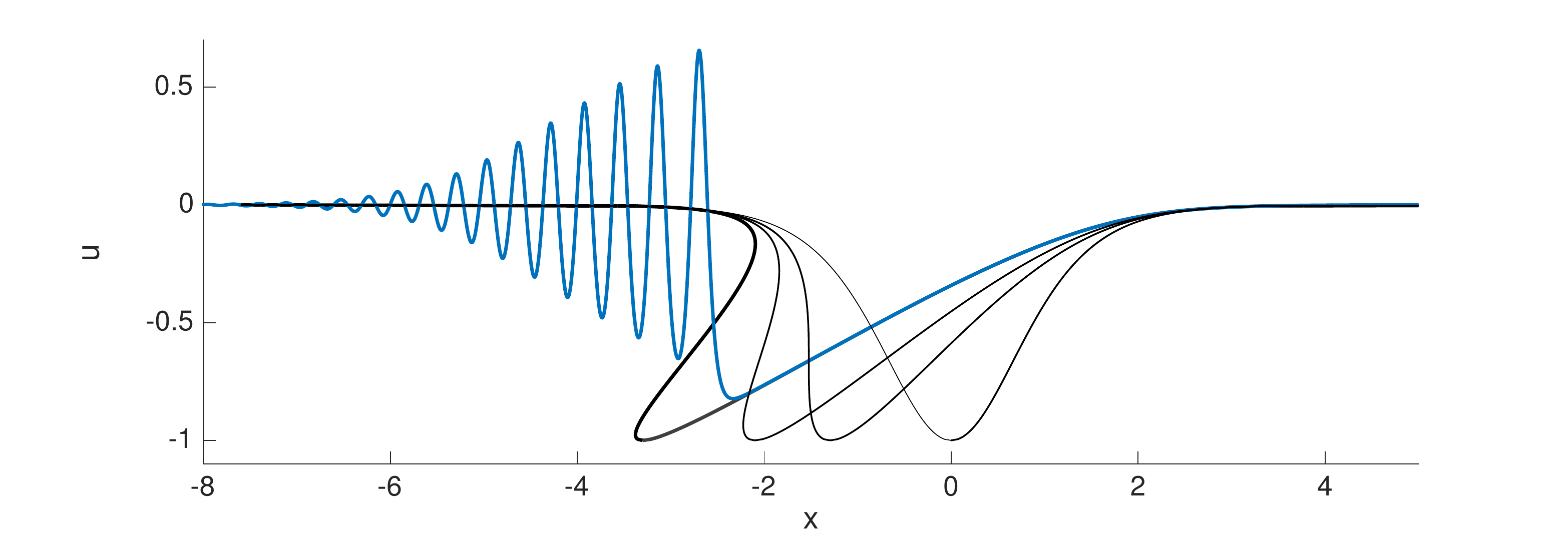}
 \caption{ In blue the solution of the KdV equation for  the initial data  $f(x)=-{\rm sech}^2(x)$  at the time $t=0.55$ for $\epsilon=10^{-1}$. In black the (multivalued) solution of the Hopf equation  for the same initial data and  for several times:
 $t=0$, $t=t_c=0.128$, $t=0.35$ and $t=0.55$.  }
 \label{fig1}
\end{figure}

The  main  steps for the description of the dispersive shock waves  are   the following:
\begin{itemize}
\item as long as the solution of the Cauchy problem for  Hopf equation $v_t+6vv_x=0$ with the  initial data $v(x,0)=f(x)$ exists, then the solution of the KdV equation $u(x,t;\epsilon)=  v(x,t)+O(\epsilon^2)$.
Generically the solution of the Hopf equation obtained by the method of characteristics
\begin{equation}
\label{Hopfsol}
v(x,t)=f(\zeta),\quad x=f(\zeta)t+\zeta,
\end{equation}
develops a singularity when the function $\zeta=\zeta(x,t)$ given implicitly by the map $x=f(\zeta)t+\zeta$ is not uniquely defined.  This happens at the first time when 
  $f'(\zeta)t+1=0$ and  $f''(\zeta)=0$ (see Figure~\ref{fig1}). These two equations and (\ref{Hopfsol}) fix uniquely the point  $(x_c,t_c)$ and $u_c=v(x_c,t_c)$. At this point,  the gradient  blow up: $v_x(x,t)|_{x_c,t_c}\to \infty$. 
 \item  The solution of the KdV equations remains smooth for all positive times.
Around the time when  the solution  of  the Hopf equation develops its first singularity at time $t_c$,  the KdV solution,  in order to compensate the formation of the strong gradient,    starts to oscillate, see Figure~\ref{fig1}.
For $t>t_c$  the solution of the KdV equation $u(x,t;\epsilon)$  is described as $\epsilon\to 0$  as follows:
\begin{itemize}
\item  there is  a cusp shape region of the $(x,t)$ plane   defined by $x_-(t)<x<x_+(t)$ with $x_-(t_c)=x_+(t_c)=x_c$. Strictly  inside the cusp, the solution    $u(x,t;\epsilon)$  has  an  oscillatory behaviour which is asymptotically described by 
the travelling wave solution (\ref{elliptic0}) where the parameters $\beta_j=\beta_j(x,t)$, $j=1,2,3$, evolve according to the Whitham 
modulation equations. 
\item Strictly outside the cusp-shape region the KdV solution is still approximated by the solution of the Hopf equation, namely $u(x,t;\epsilon)= v(x,t) +O(\epsilon^2)$. 
\end{itemize}
\end{itemize}
Later the mathematicians Lax-Levermore \cite{LL} and Venakides \cite{V1}, \cite{V2} gave a rigorous mathematical derivation  of the small dispersion limit of the KdV equation by  solving the corresponding  Cauchy problem via inverse scattering and  doing the small $\epsilon$ asymptotic.
Then  Deift, Venakides and Zhou \cite{DVZ} obtained an explicit derivation of the phase  $\phi_0$. The error term $O(\epsilon^2)$ of the expansion outside the oscillatory zone was calculated in \cite{CG4}.
For  analytic initial data,  the small $\epsilon$ asymptotic of the solution $u(x,t;\epsilon)$ of the KdV equation is given for some  times $t>t_c$  and within a  cusp
$x_-(t)<x<x_+(t)$  in the $(x,t)$ plane
 by the formula (\ref{elliptic0}) 
where  $\beta_j=\beta_j(x,t)$ solve the Whitham modulations equations (\ref{Whitham}).  The  phase $\phi_0$ in the argument of the theta-function will be described below.
In the next section we will explain how to construct the solution of the Whitham equations.
%
\subsubsection{Solution of the Whitham equations}
\begin{figure}
  \includegraphics[width=1.0\textwidth]{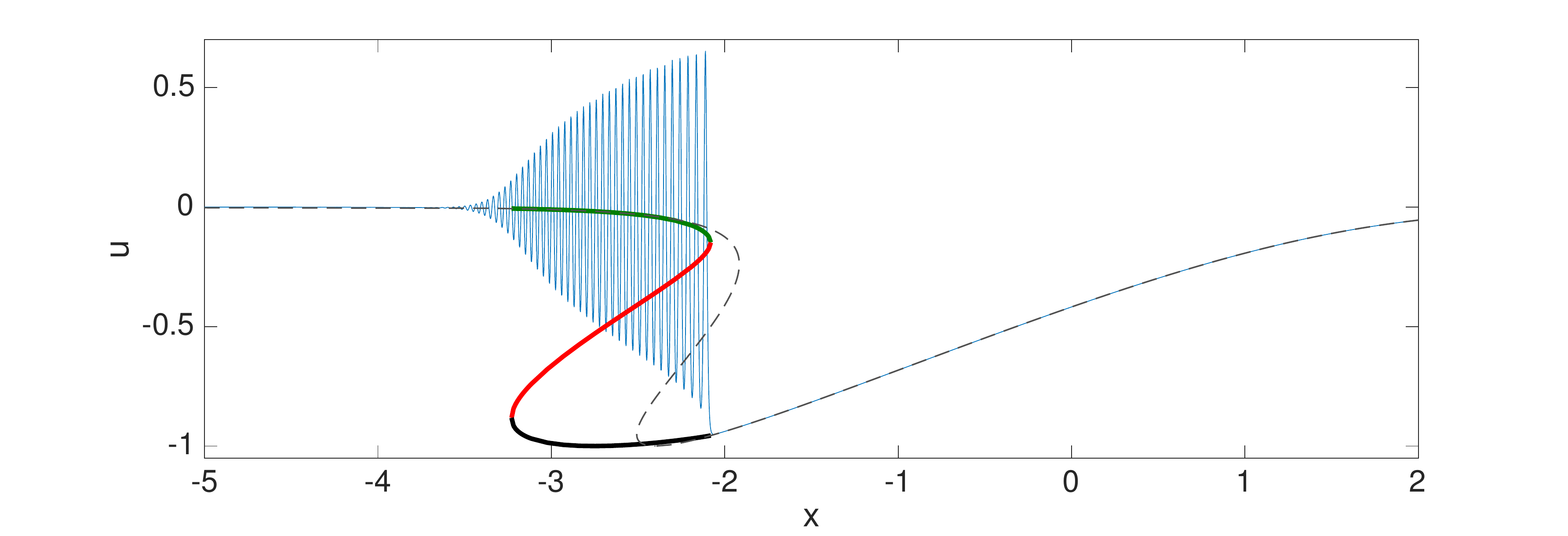}
   \caption{The thick line  (green, red and black) shows the  solution of the Whitham equations $\beta_1(x,t)\geq \beta_2(x,t)\geq \beta_3(x,t)$ at  $t=0.4$ as branches of a multivalued function  for the initial data $f(x)=-{\rm sech}^2(x)$.  At this time,  $\beta_3$ goes beyond the negative hump of the initial data and formula (\ref{qnm}) has  been used.
  The solution of the Hopf equation including the multivalued region is plotted with a dashed  grey line, while the solution   of the KdV equation  for $\epsilon=10^{-2}$ is plotted with a blue  line.
  We observe that the multivalued  region for the Hopf solution is sensible smaller then the region where the oscillations develops, while the Whitham zone is slightly smaller.
  }
 \label{KdVweak}
\end{figure}

The solution $\beta_1(x,t)>\beta_2(x,t)>\beta_3(x,t)$ of the Whitham equations can be considered as branches of a multivalued function and it is fixed by the following conditions.
\begin{itemize}
\item Let $(x_c,t_c)$ be  the  critical point where the solution of the Hopf equation develops its first  singularity and let $u_c=v(x_c,t_c)$. Then at  $t=t_c$
\[
\beta_1(x_c,t_c)=\beta_2(x_c,t_c)=\beta_3(x_c,t_c)=u_c;
\]
\item for $t>t_c$ the solution of the Whitham equations is fixed by the boundary value problem  ( see Fig.\ref{KdVweak})
\begin{itemize}
\item when $\beta_2(x,t)=\beta_3(x,t)$, then $\beta_1(x,t)=v(x,t)$;
\item when $\beta_1(x,t)=\beta_2(x,t)$,  then $\beta_3(x,t)=v(x,t)$,
\end{itemize}
where $v(x,t)$ solve the Hopf equation.   
\end{itemize}
From   the integrability of the KdV equation,  one has the integrability of the Whitham equations \cite{DN}.  This is a non trivial fact. However we give it for granted and assume that the Whitham equations have an infinite family of commuting flows:
\[
\dfrac{\partial}{\partial s}\beta_i+w_i\dfrac{\partial}{\partial x}\beta_i=0,\quad i=1,2,3.
\]
The compatibility condition of the above flows  with the Whitham equations (\ref{Whitham}),  implies that   $\dfrac{\partial}{\partial t}\dfrac{\partial}{\partial s}\beta_i=\dfrac{\partial}{\partial s}\dfrac{\partial}{\partial t} \beta_i$. From these compatibility conditions it follows  that 
\begin{equation}
\label{Tsarev}
 \dfrac{1}{w_i-w_j}\dfrac{\partial}{\partial \beta_j}w_i=\dfrac{1}{\lambda_i-\lambda_j}\dfrac{\partial}{\partial \beta_j}\lambda_i,\;\;i\neq j
\end{equation}
where  the  speeds $\lambda_i$'s  are defined in (\ref{Whitham}).

 Tsarev \cite{Tsarev} showed that if the  $w_i=w_i(\beta_1,\beta_2,\beta_3)$ satisfy the above linear overdetermined system,  then the formula 
\begin{equation}
\label{hodograph}
x=\lambda_it+w_i,\quad i=1,2,3,
\end{equation}
that is  a generalisation of the method of characteristics, gives a local solution of the Whitham equations (\ref{Whitham}).
Indeed by subtracting two equations in (\ref{hodograph})  with different indices we obtain
\begin{equation}
\label{t}
(\lambda_i-\lambda_j)t+w_i-w_j=0, \quad \mbox{or    } \quad t=-\dfrac{w_i-w_j}{\lambda_i-\lambda_j}.
\end{equation}
Taking the derivative with respect to   $x$ of the  hodograph equation (\ref{hodograph})  gives
\[
\sum_{j=1}^3 \left(\dfrac{\partial \lambda_i}{\partial \beta_j}t+\dfrac{\partial w_i}{\partial \beta_j}\right)\dfrac{\partial \beta_j}{\partial x}=1.
\]
Substituting in the above  formula the time as in (\ref{t}) and using (\ref{Tsarev}), one get that only the term with $j=i$ surveys, namely 
\[
\left(\dfrac{\partial \lambda_i}{\partial \beta_i}t+\dfrac{\partial w_i}{\partial \beta_i}\right)\dfrac{\partial \beta_i}{\partial x}=1.
\]
In the same way, making the derivative with respect to time of (\ref{hodograph}) one obtains
\[
\left(\dfrac{\partial \lambda_i}{\partial \beta_i}t+\dfrac{\partial w_i}{\partial \beta_i}\right)\dfrac{\partial \beta_i}{\partial t}+\lambda_i=0.
\]
The above two equations are equivalent to the Whitham system (\ref{Whitham}). The transformation (\ref{hodograph}) is called also hodograph transform. 
 To complete the integration 
 one needs to specify the quantities $w_i$ that satisfy the linear overdetermined system (\ref{Tsarev}).
As a  formal ansatz  we  look for a conservation law of the form 
\[
\partial_{s} k+ \partial_x(kq)=0,
\]
 with $k$ the wave number and  the function  $q=q(\beta_1,\beta_2,\beta_3)$  to be determined (recall that $q=2(\beta_1+\beta_2+\beta_3)$ for the Whitham equations (\ref{Whitham})). Assuming that the $\beta_i$ evolves independently,  such ansatz  gives $w_i$ of the form
\begin{equation}
    w_{i} =
    \frac{1}{2}\left(v_{i}-2\sum_{k=1}^{3}\beta_{k}\right)\frac{
    \partial q}{\partial\beta_{i}}+q,\quad i=1,2,3.
    \label{eq:w}
\end{equation}
 Plugging the  expression (\ref{eq:w})   into (\ref{Tsarev}),
 one obtains equations  for the function $q=q(\beta_1,\beta_2,\beta_3)$
\begin{equation}
\label{eq_q}
\dfrac{\partial q}{\partial \beta_i}-\dfrac{\partial q}{\partial \beta_j}=2(\beta_i-\beta_j)\dfrac{\partial^2 q}{\partial \beta_i\partial \beta_j},i\neq j,\;i,j=1,2,3.
\end{equation}
Such system of equations  is a linear over-determined system of Euler-Poisson Darboux type and it was obtained in \cite{GKE} and \cite{FRT2}.  The boundary conditions on the $\beta_i$ specified at the beginning of the section fix uniquely the solution.
The integration  of (\ref{eq_q}) was performed for particular initial data in several different works (see e.g.  \cite{Kamchatnov}, or \cite{Novikov}, \cite{GKE}) and for general smooth  initial data in   \cite{FRT2},\cite{FRT3}. The boundary conditions  require that  when 
$\beta_1=\beta_2=\beta_3=\beta$, then $q(\beta,\beta,\beta)=h_L(\beta)$ where $h_L$ is the inverse of the decreasing part of the initial data $f(x)$. The resulting function $ q(\beta_1,\beta_2,\beta_3)$ is \cite{FRT2}
\begin{equation}
\label{q0}
    q(\beta_{1},\beta_{2},\beta_{3}) = \frac{1}{2\sqrt{2}\pi}
    \int_{-1}^{1}\int_{-1}^{1}d\mu d\nu \frac {h_L( \frac{1+\mu}{2}(\frac{1+\nu}{2}\beta_{1}
	+\frac{1-\nu}{2}\beta_{2})+\frac{1-\mu}{2}\beta_{3})}{\sqrt{1-\mu}
    \sqrt{1-\nu^{2}}}.
\end{equation}
For initial data with a single negative hump,   such formula is valid as long as $\beta_3>f_{min}$ which is the minimum value of the initial data. When $\beta_3$ goes  beyond the hump one needs to take into account also   the increasing part $h_R$ of the inverse the initial data $f$, namely  \cite{FRT3}
\begin{equation}
\label{qnm}
q(\beta_1,\beta_2,\beta_3)=\frac{1}{2\pi}\int_{\beta_2}^{\beta_1} \dfrac{d\lambda\left(\displaystyle\int_{\beta_{3}}^{-1}
    \frac{d\xi
    h_{R}(\xi)}{\sqrt{\lambda-\xi}}+\displaystyle\int_{-1}^{\lambda}
    \frac{d\xi h_{L}(\xi)}{\sqrt{\lambda-\xi}}\right)
}{
\sqrt{(\beta_1-\lambda)(\lambda-\beta_2)(\lambda-\beta_3)}}.
\end{equation}

%
%
%
%
%
Equations (\ref{hodograph}) define $\beta_j$, $j=1,2,3$, in an implicit way as a function of $x$ and $t$. The actual solvability of (\ref{hodograph}) for $\beta_j=\beta_j(x,t)$
was obtained in a series of papers by Fei-Ran Tian \cite{FRT1} \cite{FRT3}  (see Fig.~\ref{KdVweak}). 
The Whitham equations are a systems of hyperbolic equations, and generically their  solution can suffer  blow up of the gradients in finite time.
When this happen  the small $\epsilon$ asymptotic of the solution of the KdV equation is described by higher order $\theta$-functions and the so called multi-phase Whitham equations \cite{FFM}.
 So generically speaking the solvability  of system (\ref{hodograph}) is not an obvious fact.
 The main results of \cite{FRT1},\cite{FRT2} concerning this issue are the following:
\begin{itemize}
\item  if the decreasing part of the initial data, $h_L$ is such that $h'''_L(u_c)<0$  (generic condition) then the solution of the Whitham equation exists for short times $t>t_c$.
\item  If  furthermore, the initial data $f(x)$ is step-like and non increasing,    then under some mild extra assumptions,
 the solution of the Whitham equations exists for short times $t>t_c$ and  for all times $t>T$ where $T$ is a sufficiently large time.
\end{itemize}
These results show that  the Gurevich Pitaevski description of the dispersive shock waves is generically  valid for short times $t>t_c$ and,
  for non increasing initial data, for  all times $t>T$ where $T$ is  sufficiently large.
At the intermediate times, the asymptotic description  of the KdV solution is generically given  by the modulated multiphase solution of KdV (quasi-periodic in $x$ and $t$ )  where the wave parameters evolve according to the multi-phase  Whitham equations \cite{FFM}. The study of these intermediate times  has been considered in  \cite{GT}, \cite{ABH},\cite{AB}.

To complete the description  of the dispersive shock wave we need to specify the phase of the oscillations in (\ref{elliptic}).
 Such phase was derived in \cite{DVZ}  and takes the form
\begin{equation}
\label{phase}
\phi_0=-kq,
\end{equation}
where $k=\dfrac{\pi\sqrt{\beta_1-\beta_3}}{K(m)}$ is the wave number and the function $q=q(\beta_1,\beta_2,\beta_3)$  has been defined in (\ref{q0}) or (\ref{qnm}).
The simple  form (\ref{phase}) of the phase was obtained  in \cite{GK}.
Finally the solution of the KdV equation $u(x,t;\epsilon)$ as $\epsilon\to 0$ is described as follows
\begin{itemize}
\item in  the region strictly inside the cusp $x_-(t)<x<x_+(t)$ it is given by the asymptotic formula
\begin{equation}
\label{elliptic}
u(x,t,\epsilon)= \beta_1+\beta_2+\beta_3+2\alpha+ 2\epsilon^2\frac{\partial^2}{\partial
x^2}\log\vartheta\left(\dfrac{kx-\omega t-kq)}{2\pi \epsilon};\tau\right)+O(\epsilon)
\end{equation}
where  $\beta_j=\beta_j(x,t)$ is the solution of the  Whitham  equation constructed in this section.
The wave number $k$, the frequency $\omega$ and the quantities  $\tau$ and $\alpha$ are defined in (\ref{Omega}), (\ref{tau}) and (\ref{alpha}) respectively and $q$ is defined in (\ref{q0}) and (\ref{qnm}).  
When performing the $x$-derivative in (\ref{elliptic}) observe that 
\[
\partial_x(kx-\omega t-kq)=k,
\]
 because of (\ref{hodograph}) and (\ref{eq:w}).
 \item  For $x>x_+(t)+\delta$ and $x<x_-(t)-\delta$ for some positive $\delta>0$, the KdV solution is approximated   by
 \[
 u(x,t,\epsilon)=v(x,t)+O(\epsilon^2)
 \]
 where $v(x,t)$ is the solution of the Hopf equation.
 \end{itemize}
\begin{figure}
 \hspace{-20pt}   \includegraphics[width=0.6\textwidth]{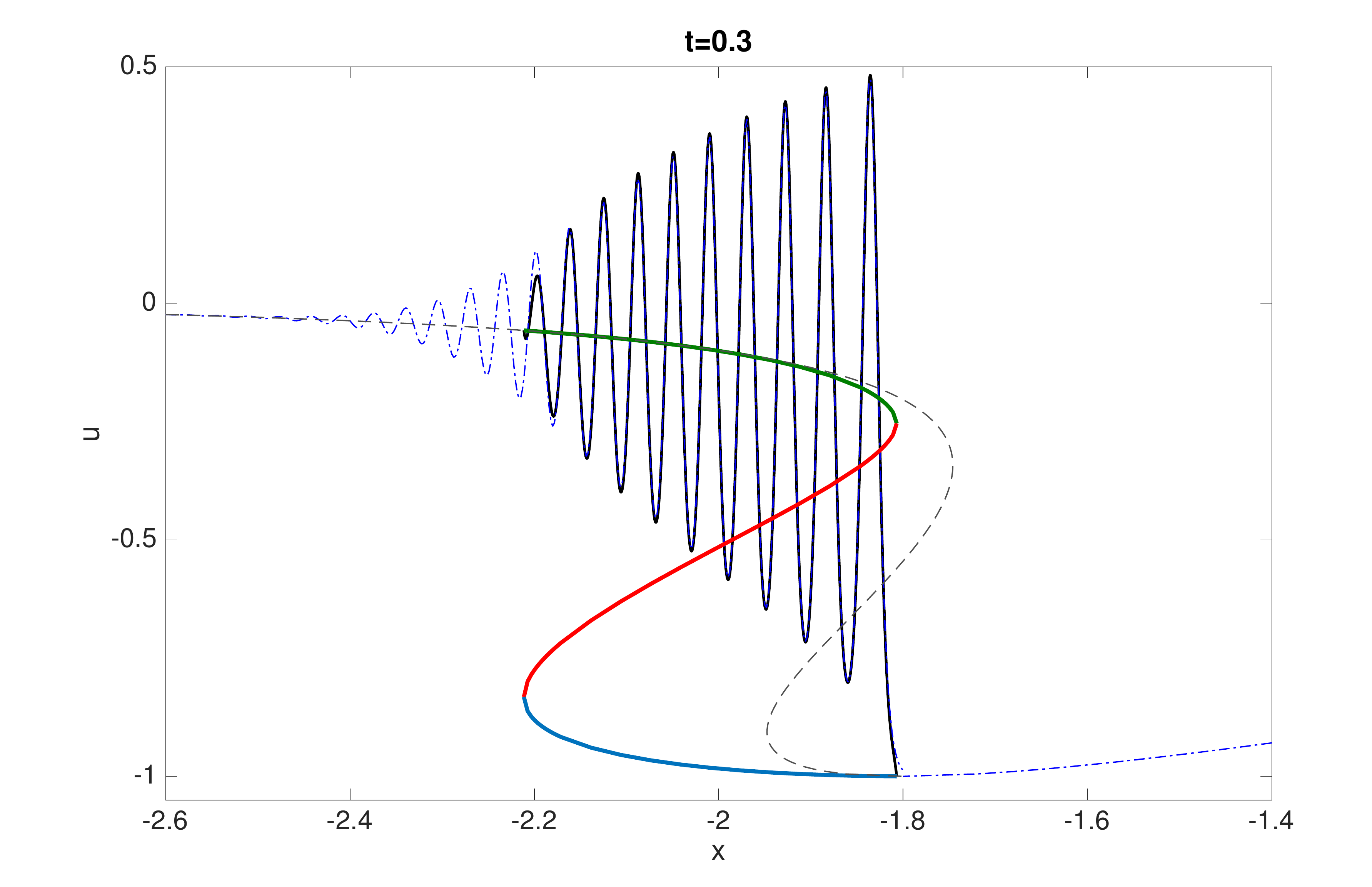}
  \hspace{-10pt} 
 \includegraphics[width=0.58\textwidth]{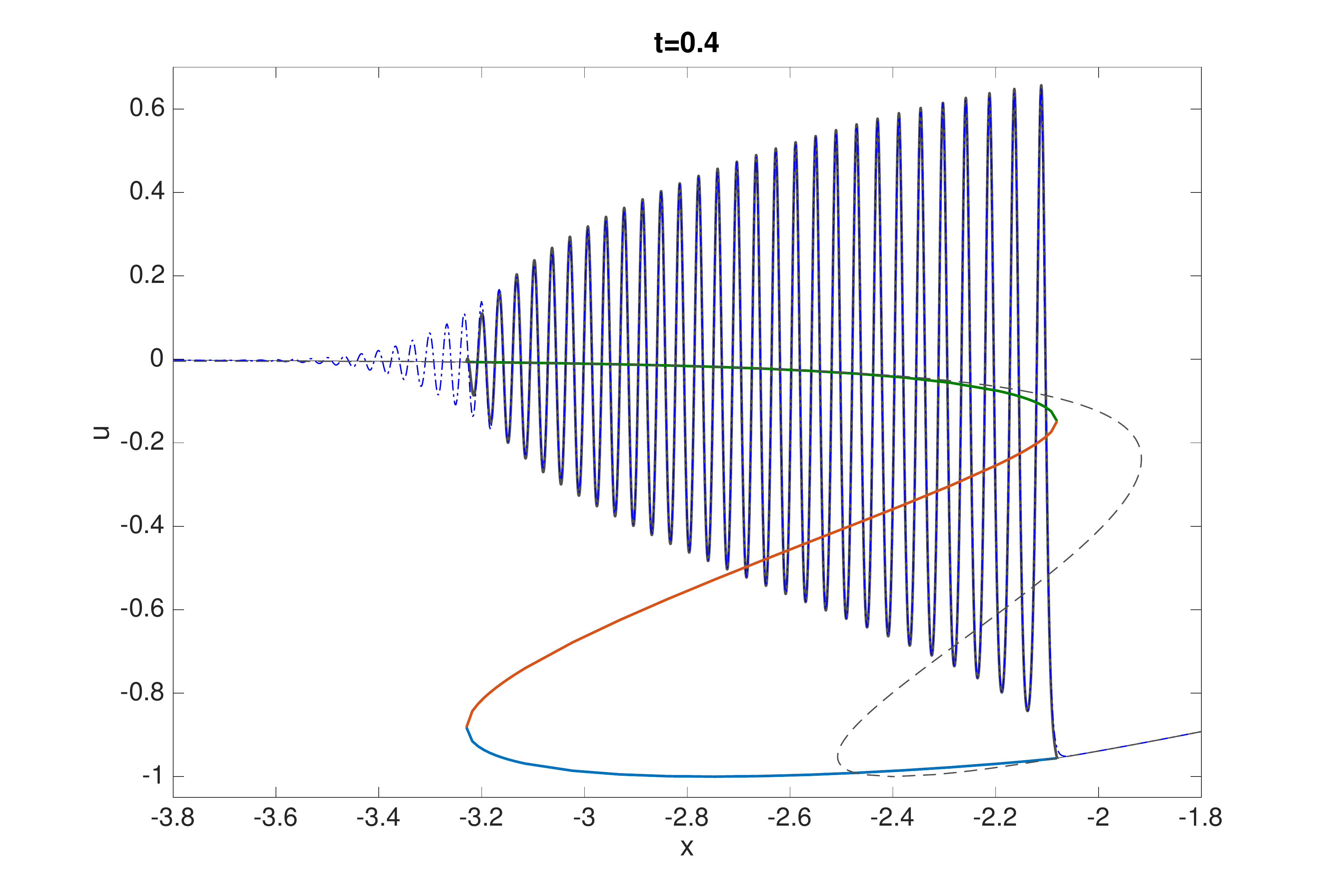}   
 \caption{ The solution of the KdV equation and its approximations  for the initial data $f(x)=-{\rm sech}^2(x)$ and $\epsilon=10^{-2}$ at two different times $t=0.3$ and $t=0.4$.
 The blu dash-dot line is the KdV solution, the black line is the elliptic asymptotic formula (\ref{elliptic}) which is on top of the KdV solution,  the  black dash line is the solution of the Hopf equation while the green, red  and  aviation  blue  lines are the solution of the Whitham equations $ \beta_1\geq \beta_2\geq \beta_3$.
 }
 \label{KdVasym1}
\end{figure}

 Let us  stress the meaning of the formula (\ref{elliptic}): such formula shows that the leading order behaviour of the KdV solution $u(x,t;\epsilon)$ in the limit $\epsilon\to 0$  and for generic initial data is 
given in a cusp-shape region of the $(x,t)$ plane  by the periodic travelling wave of KdV. However to complete the description one still needs to solve an initial value problem, for three hyperbolic equations, namely the Whitham equations, but the gain is that  these  equations are independent from $\epsilon$.   

 A first approximation of the boundary  $x_{\pm}(t)$ of the oscillatory zone for $t-t_c$ small, has been obtained in \cite{GK} by taking the limit of (\ref{hodograph}) when $\beta_1=\beta_2$ and $\beta_2=\beta_3$.
This gives
\begin{align*}
&x_+(t)\simeq x_c+6u_c(t-t_c)+\dfrac{4\sqrt{10}}{3\sqrt{-h_L'''(u_c)}}(t-t_c)^{\frac{3}{2}}, \\
&x_-(t)\simeq x_c+6u_c(t-t_c)-\dfrac{36\sqrt{2}}{\sqrt{-h_L'''(u_c)}}(t-t_c)^{\frac{3}{2}},
\end{align*}
where $h_L$ is the decreasing part of the initial data. Such formulas coincide with the one obtained in \cite{GP1} for cubic initial data.

We conclude pointing out that in \cite{GK} a  numerical comparison of the asymptotic formula (\ref{elliptic}) with the actual KdV solution $u(x,t;\epsilon)$ has been considered for the intial data $f(x)=-\mbox{sech}^2x$.
Such numerical comparison has shown  the existence of transition zones  between the oscillatory  and non oscillatory regions  that are described by Painlev\'e trascendant and elementary functions \cite{CG1},\cite{CG2},\cite{CG3}. Looking for example to  Fig. \ref{KdVasym1} it is clear that the KdV oscillatory region  is slightly larger then the region described  by the elliptic asymptotic 
(\ref{elliptic}) where the oscillations are confined to  $x_-(t)\leq x\leq x_+(t)$.

Of particular interest is the solution of the KdV equation near the region where the oscillations are almost linear, namely near the point $x_-(t)$.
 It is known   \cite{FRT1, GT}  that taking the limit of the hodograph transform (\ref{hodograph})  when  $\beta_2= \beta_3=\xi$  and $\beta_1=v$,   one obtains the system of equations
 \begin{eqnarray}
  \label{sys_lead}
\left\{ \begin{array}{ll}
&\label{leading}x_-(t)=6tv(t)+h_L(v(t)),\\
&\label{leading2}6t+\phi(\xi(t);v(t))=0,\\
&\label{leading3}\partial_{\xi}\phi(\xi(t);v(t))=0,
\end{array}\right.
\end{eqnarray}
 that determines uniquely $x_-(t)$ and and  $v(t)>\xi(t)$. In the above equation the function 
\begin{equation}
\label{theta}
\phi(\xi;v)=\dfrac{1}{2\sqrt{v-\xi}}\int_{\xi}^v\dfrac{h'_L(y)dy}{\sqrt{y-\xi}},
\end{equation}
and $h_L$ is the decreasing part of the initial data.
The behaviour of the KdV solution is described  near the edge $x_-(t)$ by  linear oscillations, where the envelope of the oscillations is given by  the Hasting Mcleod solution to the Painlev\'e II equation:
 \begin{equation}\label{PII}
 q''(s)=sq+2q^{3}(s).
\end{equation}
The special solution in which we are interested, is the Hastings-McLeod solution \cite{HastingsMcLeod}
which is uniquely determined by the boundary conditions
\begin{align}
&\label{HM1}q(s)=\sqrt{-s/2}(1+o(1)),&\mbox{ as $s\to -\infty$,}\\
&\label{HM2}q(s)=\mbox{Ai}(s)(1+o(1)), &\mbox{ as $s\to +\infty$,}
\end{align}
where $\mbox{Ai}(s)$ is the Airy function.
Although any Painlev\'e II solution has an infinite number of poles in the complex plane, the Hastings-McLeod solution $q(s)$ is smooth for all real values of $s$ \cite{HastingsMcLeod} .

The KdV  solution  near $x_-(t)$ and in the limit  $\epsilon\to0$ in such a way that 
\begin{equation}
\nonumber
\lim\limits_{\substack{\epsilon\to 0  \\ x\to x_-(t)}} \dfrac{x- x^-(t)}{\epsilon^{2/3}},
\end{equation}
remains finite, is given by  \cite{CG2}
\begin{equation}
\label{expansionu}
u(x,t,\epsilon) = v(t)-\dfrac{4\epsilon^{1/3}}{c^{1/3}}q\left(s(x,t,\epsilon)\right)
\cos\left(\frac{\Theta(x,t)}{\epsilon}\right)+O(\epsilon^{\frac{2}{3}}).
\end{equation}
where 
\begin{equation}
\nonumber
\Theta(x,t)=2\sqrt{v-\xi}(x-x^-)+2\int_{\xi}^{v}(h_L'(y)+6t)\sqrt{y-\xi}dy
\end{equation}
and 
\begin{equation}\nonumber
c=-\sqrt{v-\xi}\dfrac{\partial^2}{\partial \xi^2}\phi(\xi;v)>0,\qquad s(x,t,\epsilon)=-\frac{x-x_-(t)}{c^{1/3}\sqrt{v-\xi}\,\epsilon^{2/3}}.
\end{equation}

\medskip

\begin{figure}
\vspace{-30pt}
  \begin{center}
     \includegraphics[width=0.6\textwidth]{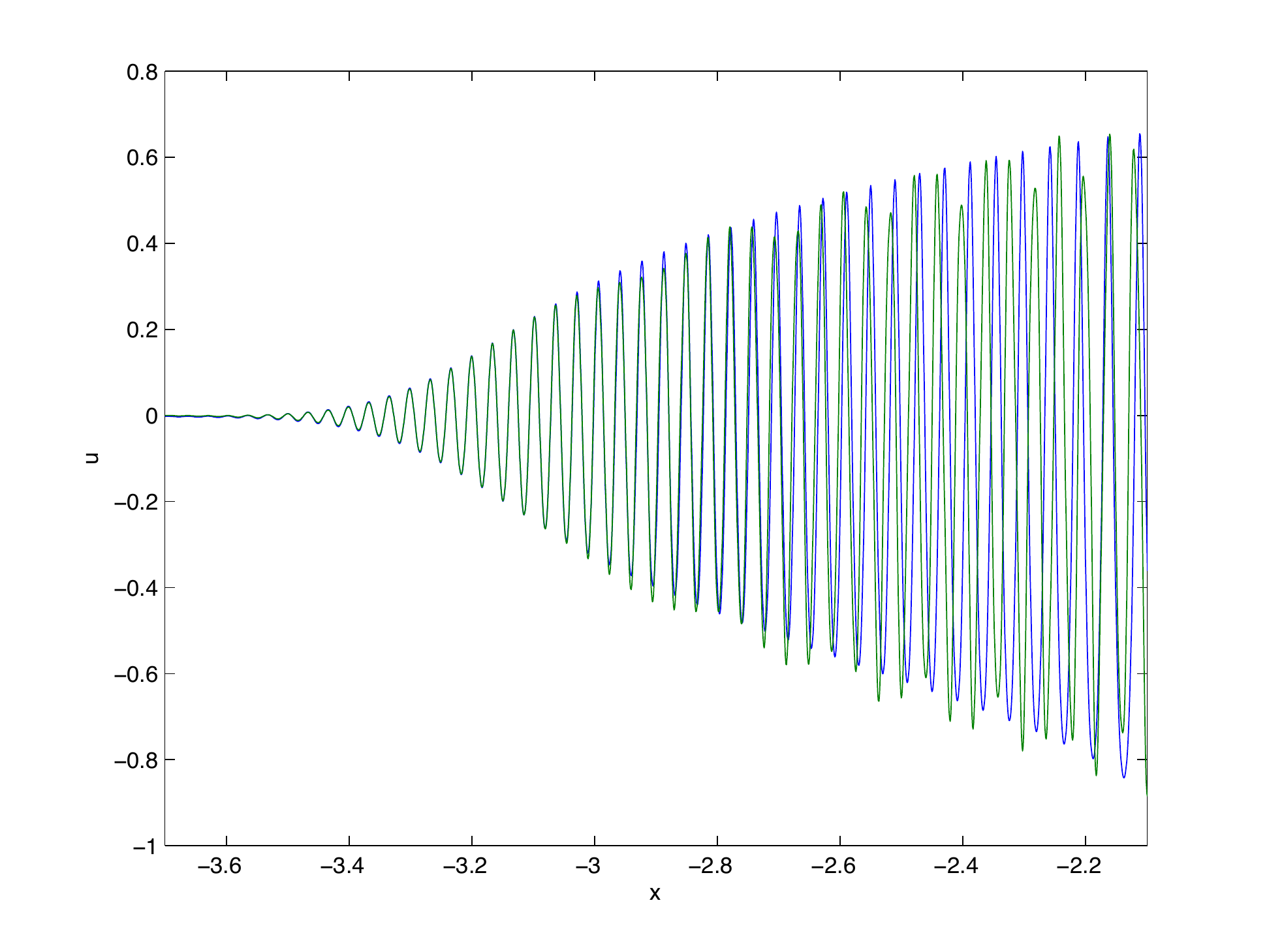}
   \caption{ The solution of the KdV equation in blue  and its approximation (\ref{expansionu})  in green for the initial data $f(x)=-{\rm sech}^2(x)$ and $\epsilon=10^{-2}$ at  $t=0.4$.
   One can see that  the green and blue lines are completely overlapped when the oscillations are small.
 }
 \label{fig_PII}
\end{center}
\end{figure}
Note that the leading order term in the expansion (\ref{expansionu}) of $u(x,t,\epsilon)$ is given by $v(t)$ that solves the Hopf equation while the oscillatory term is of order $\epsilon^{1/3}$  with oscillations of wavelength proportional to $\epsilon$ and amplitude proportional to the Hastings-McLeod solution $q$ of the Painlev\'e II equation.
From the practical point of view it is  easier to use formula (\ref{expansionu}), then (\ref{elliptic}) 
 since one needs to solve only an ODE (the Painlev\'e II equation) and three algebraic equations, namely (\ref{sys_lead}). One can see from figure (\ref{fig_PII}) that the asymptotic formula (\ref{expansionu})
 gives a  good  approximation (up to an error $O(\epsilon^{\frac{2}{3}})$) of the KdV solution near  the leading edge where the oscillations are linear, while inside the Whitham zone, it gives a  qualitative description of the oscillations \cite{GK2}.

Another interesting asymptotic regime is obtained when one wants to describe the first few oscillations of the KdV solution  in the small dispersion limit. In this case the so called  Painlev\'e I2  asymptotics  should be used. Furthermore we point out that it is simpler to solve one ODE, rather then the Whitham equations.
For example, near the critical point $x_c$ and near the critical time the following asymptotic behaviour has been conjectured in \cite{D1} and  proved in \cite{CG1}
\begin{equation}
\label{univer}
u(x,t,\epsilon)\simeq u_c +\left(\dfrac{2\epsilon^2}{\sigma^2}\right)^{1/7}
U \left(
\dfrac{x- x_c-6u_c (t-t_c)}{(8\sigma\epsilon^6)^{\frac{1}{7}}};
\dfrac{6(t-t_c)}{(4\sigma^3\epsilon^4)^{\frac{1}{7}}}\right) +O\left( \epsilon^{4/7}\right),
\end{equation}
where $\sigma=-h_L'''(u_c)$,
and $U=U(X,T)$ is the unique
real smooth solution to the fourth order ODE  \cite{CV}
\begin{equation}\label{PI2}
X=T\, U -\left[ \dfrac{U^3}{6}  +\dfrac{1}{24}( U_{X}^2 + 2 U\, U_{XX} )
+\frac1{240} U_{XXXX}\right],
\end{equation}
which is  the second member of the Painlev\'e I hierarchy (PI2 ). The relevant solution  is uniquely \cite{CL}
characterized by the asymptotic behavior 
     \begin{equation}
\label{PI2asym}
        U(X,T)=\mp (6|X|)^{1/3}\mp \frac{1}{3}6^{2/3}T|X|^{-1/3}
            + O(|X|^{-1}),
            \qquad\mbox{as $X\to\pm\infty$,}
\end{equation}
for each fixed $T\in\mathbb{R}$.  Such  Painlev\'e solution matches,  the elliptic solution (\ref{elliptic})  for the cubic inital data $f(x)=-x^{\frac{1}{3}}$ for large times \cite{C1}.
Such solution of the PI2 has been conjectured to describe the initial time  of  the formation of dispersive shock waves for general Hamiltonian perturbation of hyperbolic equations \cite{DGKM}.

 We conclude by stressing that   the  asymptotic descriptions  reviewed in this chapter for the KdV equation  can be developed for other integrable equations like the nonlinear Schr\"odinger equation, \cite{KMM} the Camass-Holm equation \cite{BIS}  or the modified KdV equation \cite{KM}. 

\section*{Acknowledgements} 
 T.G. acknowledges the support by the Leverhulme Trust Research Fellowship RF-2015-442 from UK and  PRIN  Grant ÒGeometric and analytic theory of Hamiltonian systems in finite and infinite dimensionsÓ of Italian Ministry of Universities and Researches.

\end{document}